# How geodesy can contribute to the understanding and prediction of earthquakes

G. F. Panza[1,2,3], A. Peresan[3,4], F. Sansò[2,5], M. Crespi[6], A. Mazzoni[6], A. Nascetti[6]

[1]Institute of Geophysics, China Earthquake Administration, Beijing, China
[2]Accademia Nazionale dei Lincei, Rome, Italy
[3]International Seismic Safety Organization (ISSO) - www.issoquake.org
[4]CRS – Istituto Nazionale di Oceanografia e Geofisica Sperimentale, Trieste, Italy
[5]DICA - Politecnico di Milano, Milan, Italy
[6]Geodesy and Geomatics Division – DICEA – University of Rome "La Sapienza", Rome, Italy

**Abstract**

Earthquakes cannot be predicted with precision, but algorithms exist for intermediate-term middle range prediction of main shocks above a pre-assigned threshold, based on seismicity patterns. Few years ago, a first attempt was made in the framework of project SISMA, funded by Italian Space Agency, to jointly use seismological tools, like CN algorithm and scenario earthquakes, and geodetic methods and techniques, like GPS and SAR monitoring, in order to effectively constrain priority areas where to concentrate prevention and seismic risk mitigation.
We present a further development of integration of seismological and geodetic information, clearly showing the contribution of geodesy to the understanding and prediction of earthquakes.
As a relevant application, the seismic crisis that started in Central Italy in August 2016 with the Amatrice earthquake and still going on, is considered in a retrospective analysis of both GPS and SAR data.
Differently from the much more common approach, here GPS data are not used to estimate the standard 2D velocity and strain field in the area, but to reconstruct the velocity and strain pattern along transects, which are properly oriented according to the a priori information about the known tectonic setting. SAR data related to the Amatrice earthquake coseismic displacements are here used as independent check of the GPS results.
Overall, the analysis of the available geodetic data indicates that it is possible to highlight the velocity variation and the related strain accumulation in the area of Amatrice event, within the area alarmed by CN since November 1st, 2012. The considered counter examples, across CN alarmed and not-alarmed areas, do not show any spatial acceleration localized trend, comparable to the one well defined along the Amatrice transect.
Therefore, we show that the combined analysis of the results of intermediate term middle range earthquake prediction algorithms, like CN, with those from the processing of adequately dense and permanent GNSS network data, possibly complemented by a continuous InSAR tracking, may allow the routine highlight in advance of the strain accumulation. Thus it is possible to significantly reduce the size of the CN alarmed areas.

**Introduction**

Earthquakes cannot be predicted with precision, but algorithms exist for intermediate-term middle range prediction of main shocks above a pre-assigned threshold, like M8 and CN. The alarms, which refer to areas with linear dimensions of hundred kilometres and having a duration of several months to years, are not compatible with evacuation or red alert, but can be very useful for many effective low key prevention actions (Kantorovich et al., 1974; Kantorovich and Keilis-Borok, 1991, Peresan et al., 2012). The formulation of the M8 and CN predictions satisfies the basic principle of science introduced by Karl Popper: a model to be scientific acceptable must be falsifiable. The statistical validity has been proven both at global and regional scale (Kossobokov, 2014; Kossobokov and Soloviev, 2015). Indeed "The proof of the pudding is in the eating…"

CN prediction experiment started in Italy in 1998 (Peresan et al., 2005). Since then nine strong earthquakes occurred in the territory monitored by CN and seven of them have been predicted in "real time" – the event occurred after the alarm was declared to a group of scientists and administrators – more than one hundred people - who routinely receive prediction results (Peresan, 2017).

At the meeting of the Commissione Grandi Rischi (CGR) of May 4th 2012, the reliability of CN alarm for "Northern (Italy) region" has been questioned, but the May 21st 2012 earthquake in Emilia tragically confirmed the alarm. Similar predictions have been made before the earthquakes of Amatrice and Norcia (DMG, 2017; IEPT, 2017).

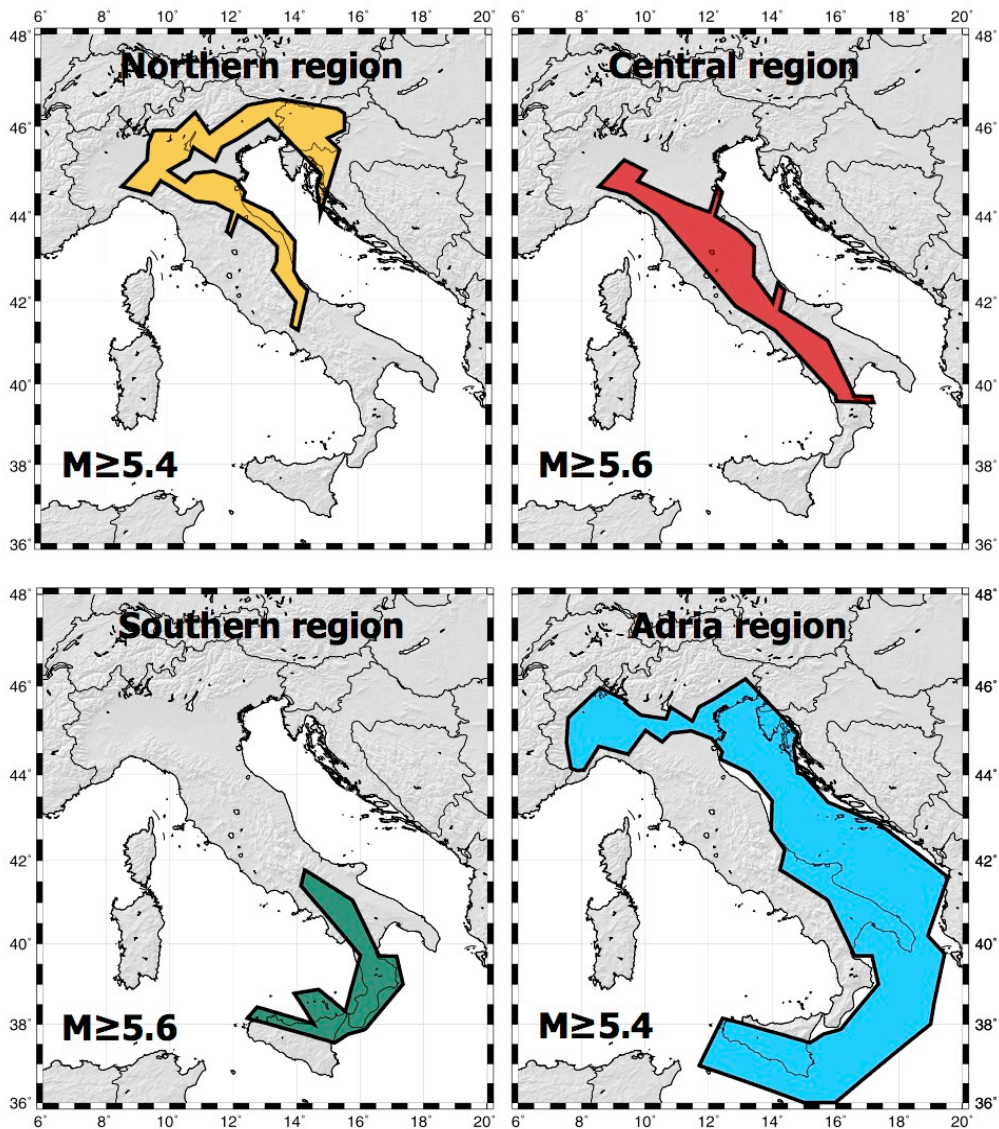

Figure 1 - Regionalization used in the CN prediction experiment in Italy (Peresan, 2017 and references therein). The regionalization is composed by four partially overlapping regions, defined based on the seismotectonic model. The magnitude threshold identifying the target earthquakes is given for each region.

The probability gain of CN predictions is around 3, i.e. the occurrence probability of a target earthquake increases by a factor of 3 during the time interval occupied by the alarm, with respect to "normal conditions" (Peresan et al., 2015). For example, in the "Northern (Italy) region" as defined by CN (see fig. 1) the probability of a target earthquake occurrence within one year, increases from 15% (average probability in normal conditions, with no predictive information) to 48% (during alarms). Similarly, in "Central region" during an alarm the probability of a target earthquake increases by about a factor 4 (from about 12% to 45%). Given the statistic achieved so far by CN algorithm, the probability of observing a target earthquake when no alarm is declared within the region is pretty small, i.e. about 2-3%. Thus, although being characterized by uncertainties at the intermediate space-time scale, CN predictions are robust and significant, as they are validated by the experiment started in 1998 and provide not negligible probabilities of strong earthquake occurrence. A summary of the CN predictions since 1998 is given in Table 1.

Table 1 – Summary of CN prediction in Italy since 1998

| Date | Latitude, °N | Longitude, °E | Depth, km | $M_{prio}$ | CN | CN Region |
|---|---|---|---|---|---|---|
| 1998.04.12 | 46.24 | 13.65 | 10 | 6.0 | Yes | North |
| 1998.09.09 | 40.03 | 15.98 | 10 | 5.7 | Yes | Centre, South |
| 2003.09.14 | 44.33 | 11.45 | 10 | 5.6 | Yes | North |
| 2004.07.12 | 46.30 | 13.64 | 7 | 5.7 | Yes | North |
| 2004.11.24 | 45.63 | 10.56 | 17 | 5.5 | No | North |
| 2009.04.06 | 42.33 | 13.33 | 8 | 6.3 | No | Centre |
| 2012.05.20 | 44.90 | 11.23 | 6 | 6.1 | Yes | North |
| 2016.08.24 | 42.72 | 13.19 | 4 | 6.2 | Yes | Centre |
| 2016.10.30 | 42.85 | 13.09 | 10 | 6.6 | Yes | Centre |

Few years ago Panza et al. (2011) in the framework of a multidisciplinary investigation supported by ASI – the project SISMA – have shown that the joint use of seismological tools, like CN algorithm and scenario earthquakes, and of geodetic methods and techniques, like GPS and SAR monitoring, permits to effectively reduce the areas where to concentrate prevention and seismic risk mitigation.

In this paper we present a further development of the integration of seismological and geodetic information, highlighting the contribution of geodesy to the understanding and prediction of earthquakes. As relevant application, we consider the ongoing seismic crisis, which is shattering Central Italy since August 2016, to exemplify the possibilities of the integrated analysis.

**The case of Amatrice earthquake**

It is well known since at least three decades that GPS (now GNSS) and SAR can provide valuable information about the ground (and, at the largest scale, crustal) deformation on a continuously increasing space and time basis. Anyway, a still open question is if this information can be useful to better understand the earthquake preparation process and to contribute to improve the assessment of the seismic hazard even at local scale.

In fact, under the hypothesis that earthquakes occur where strain energy has accumulated, space geodesy techniques can give such a contribution if they are able to reliably estimate the accumulation of strain, and if proper geophysical models are established that relate short-term strain estimation to long-term fault activity (Wright, 2016), complementing the space- and time-dependent information provided by seismic flow monitoring, e.g. the pilot project in Italy, Panza et al. (2011) and allowing the identification of critical seismogenic areas. In this respect, data from about 20000 GNSS stations have been analysed and a global map of tectonic strain has been produced and publicly available is (Kreemer et al., 2014). Moreover, based on a posteriori analysis of SAR data (Wright et al., 2013), time-dependent deformation during the earthquake preparation process has been widely detected.

However, there are large gaps in the observations coverage in many areas. Even in the most developed countries, where there are the densest GNSS networks (Japan, USA), stations are spaced between 10 and 50 km, still generally too far apart to clearly distinguish between locked faults accumulating strain and those that creep steadily.

This is even more difficult in transition zones between tectonic plates, as the Italian Apennines, where there is a system of nearly parallel normal faults, each of which can be locked and prone to accumulate strain or not. Amatrice earthquake (August 24, 2016, M=6.0) just occurred in this complicated tectonic setting (Doglioni and Panza, 2015).

The available geodetic data analysed in this work are from GPS (Devoti et al., 2016) and from SAR (ascending and descending mode interferograms from Sentinel-1A and 1B, respectively related to August 15 and 21, 2016, before the event, and August 27, 2016, after the event) (Piccardi et al., 2016).

The overall goal pursued using these data is to understand whether it would have been possible to highlight the strain accumulation related to the preparation of the Amatrice earthquake using GPS data, and using SAR data as an independent check of the GPS results.

Accordingly with Marcel Proust: "The real voyage of discovery consists not in seeking new lands but in seeing with new eyes". Therefore, in this study the analysis of the GPS data is not devoted, as it is much more common, to estimate the 2D velocity and strain field in the area. Rather, benefiting from the CN regionalization and from the known tectonic setting as a priori information, the analysis aims to reconstruct the velocity and strain pattern along a transect, 50 km wide, orthogonal to the known fault system. This width has been chosen as reasonable compromise between the opposite needs to consider a substantial number of GPS stations and to focus on an area with a homogeneous, as much as possible, tectonic setting. The area resulting from the intersection between the outlined transect and the CN Central Region is about 200 km long (from the Tyrrhenian to the Adriatic Sea coasts) and crosses the Apennines along the direction of maximum tectonic extension (approximate azimuth 55°), with the axis passing through Amatrice (Figure 2). In the following this area is referred as "Amatrice area".

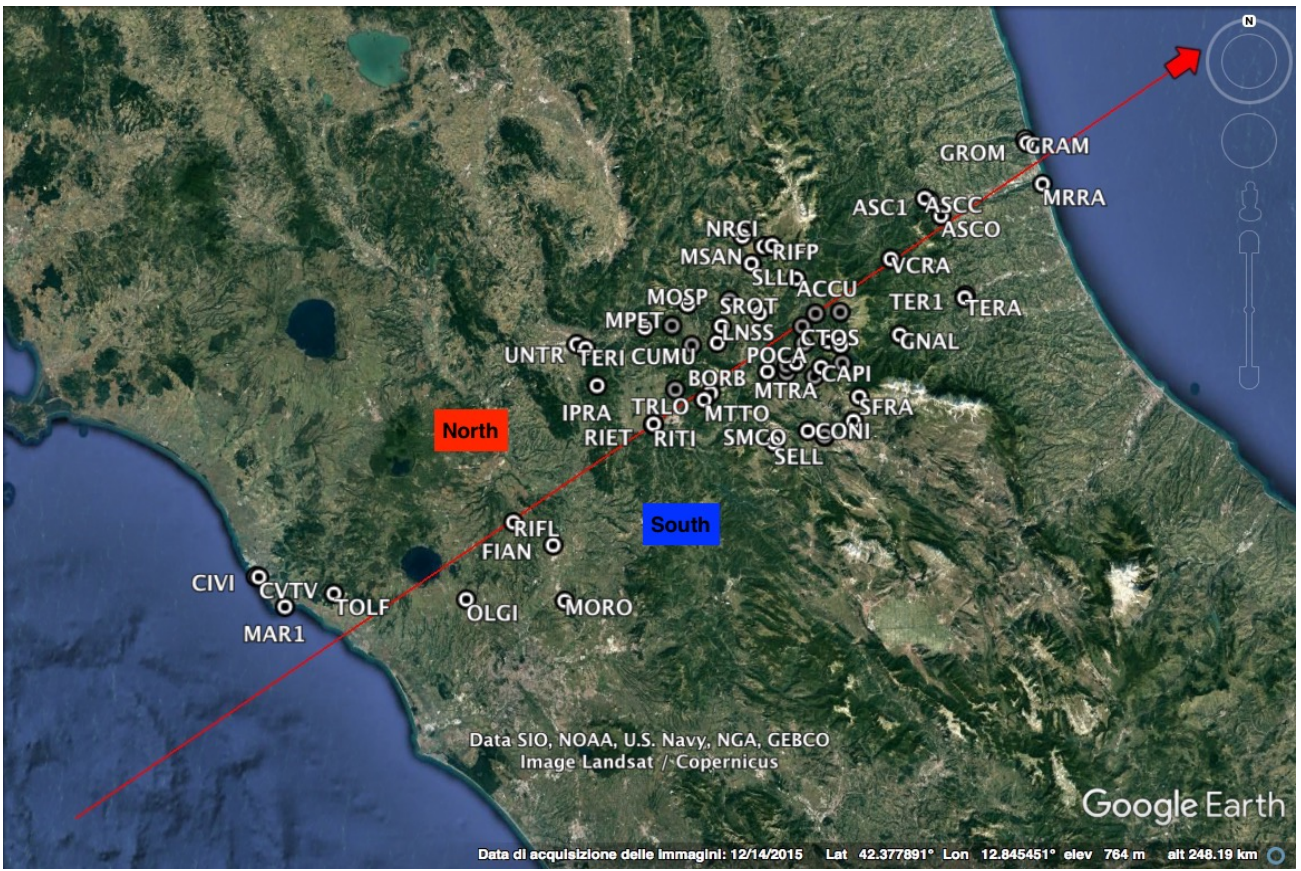

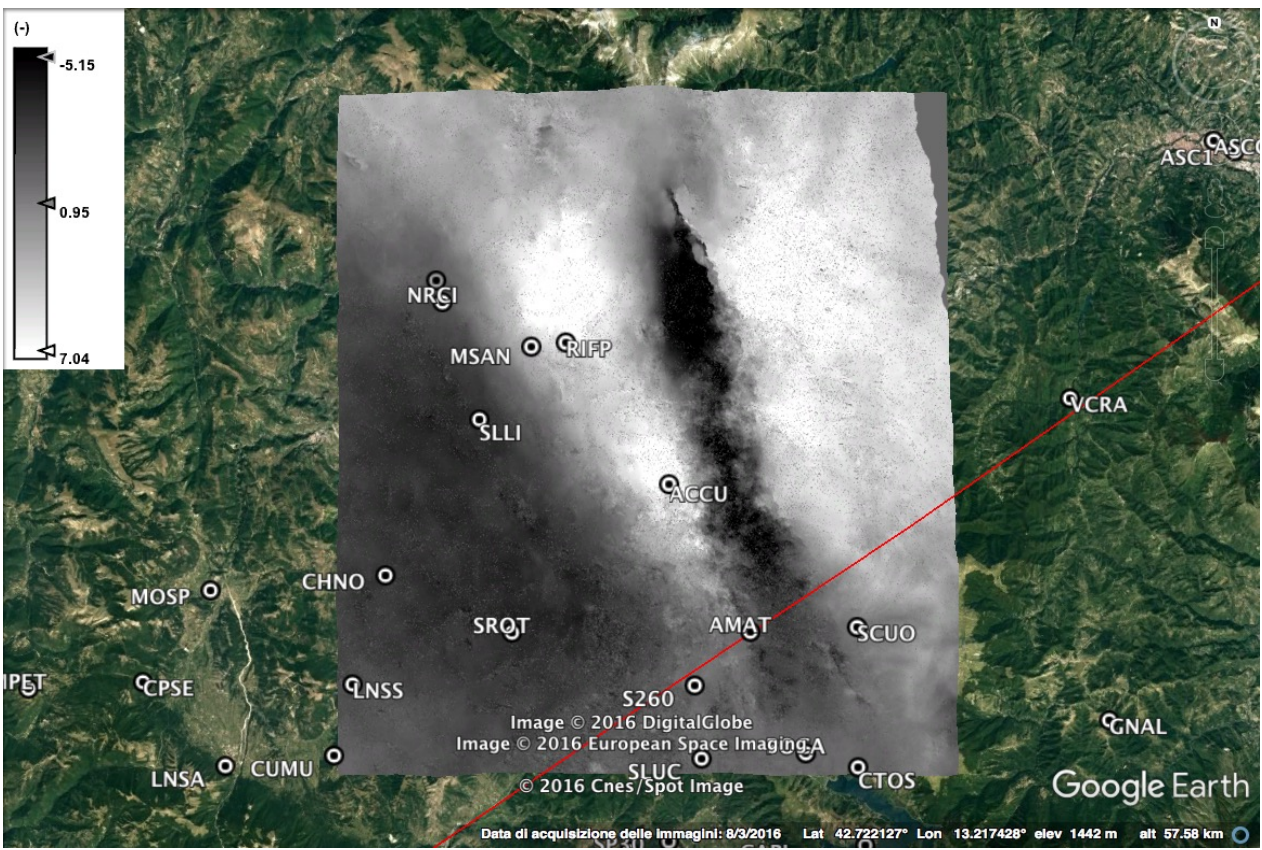

Figure 2 - Amatrice transect across the Apennines with GNSS stations (above), and (below) zoom on Amatrice area (with West-East coseismic displacement field from SAR – in cm, positive eastward, see Figure 4).

At first, GPS data, approximately related to the period beginning of January 2005 – middle of August 2016 - just before the Amatrice earthquake - but with a significant dispersion (this dispersion is due to the standard evolution of the geodetic monitoring network, so that the indicated period represents the interval during which the most part of the GPS stations were active), evidence a clear increase of velocity in the direction of the transect (tectonic extension) moving eastward from the Tyrrhenian coast to the Amatrice area, with the peak velocity gradient in the area between Rieti and Amatrice (Figure 3: abscissa along the transect between – 50 and 0 km). This is a clear and robust indication supported by the sufficiently dense distribution of GPS stations (a large part non-permanent ones) (Galvani et al., 2013) along the transect, mainly in the area between Rieti and Amatrice (Figures 2 and 3). Moreover, it is evident the expected velocity reduction again moving eastward, between Amatrice area and the Adriatic coast, that experiences tectonic shortening.

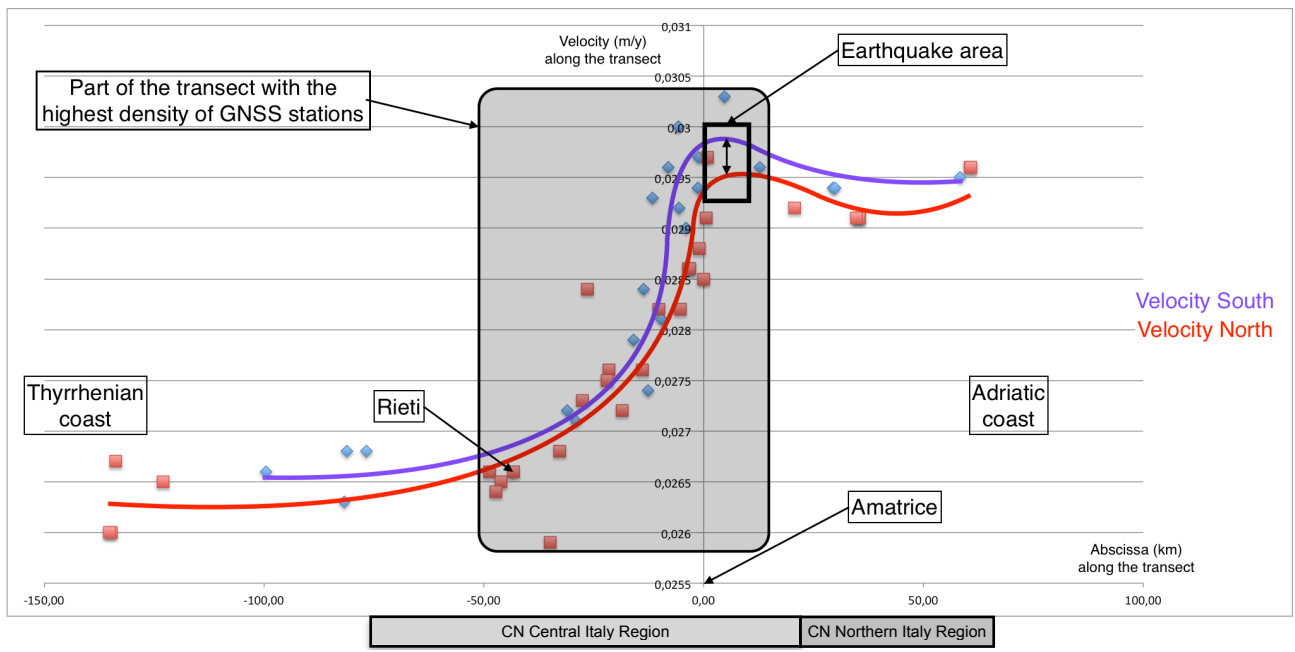

Figure 3 - Velocity pattern along the Amatrice transect across Apennines (time range: beginning of January 2005-middle of August 2016); Velocity South and North indicate the velocity patterns as derived by the GPS stations respectively located south and north with respect to the axis of the transect passing through Amatrice (see Figure 2); the intersections of the axis of the transect with the CN regions are indicated.

Furthermore, the velocity gradient in the direction of the transect is higher in the southern part of the transect than in the northern one, especially close to Amatrice, which was the area struck by the earthquake. The mean velocity difference between the southern and northern parts of the transect can be estimated (spline interpolation) around 0.25 mm/y from GPS data (Figure 3). This velocity difference, when multiplied by the mean occurrence rate of earthquakes with M≥5.6 in the above defined Amatrice area (approximately 300 years, according to Riguzzi et al., 2013), results in a displacement difference between the southern and the northern part of the transect. Incidentally $M_0$=5.6, corresponds to an average occurrence rate of about 6-7 years during the CN threshold setting time interval (1 January 1954 – 1 January 1986) in CN Central Italy (Peresan et al., 2005). The West-East component of this displacement difference (~ 0.20 mm/y x 300 y = 6 cm), which can be directly computed from the GPS data, approximately equals the westward coseismic displacement estimated from SAR (~ 6 cm) (Figure 4) in correspondence of the (locked) fault which activated during the Amatrice earthquake (INGV working group, 2016; Gruppo di Lavoro INGV, 2016; Lavecchia et al., 2016; Piccardi et al., 2016). It has to be recalled that SAR is able to estimate accurately displacements only in the plane defined by the ascending and descending line of sight that is intersecting the horizontal plane approximately in W/E direction, but in this particular situation we are lucky since this is also the approximate direction orthogonal to the activated fault (system). This particularly favourable condition approximately holds for all the Central Apennines, from Northern Marche-Umbria to Basilicata.

Therefore, there is a good coherence between the West-East component displacement difference between the southern and the northern part of the transect derived by GPS and the West-East component of the coseismic displacement in the northern part derived by SAR. This result underlines that, at the present achievable accuracy level, GPS and SAR are suited to reliably reconstruct the geodetic signatures of the two here investigated parts of the earthquake preparation process, the strain accumulation (by the GPS estimated velocity field) and the coseismic displacement field (subsequently estimated by SAR), in a small portion of the CN alarmed area.

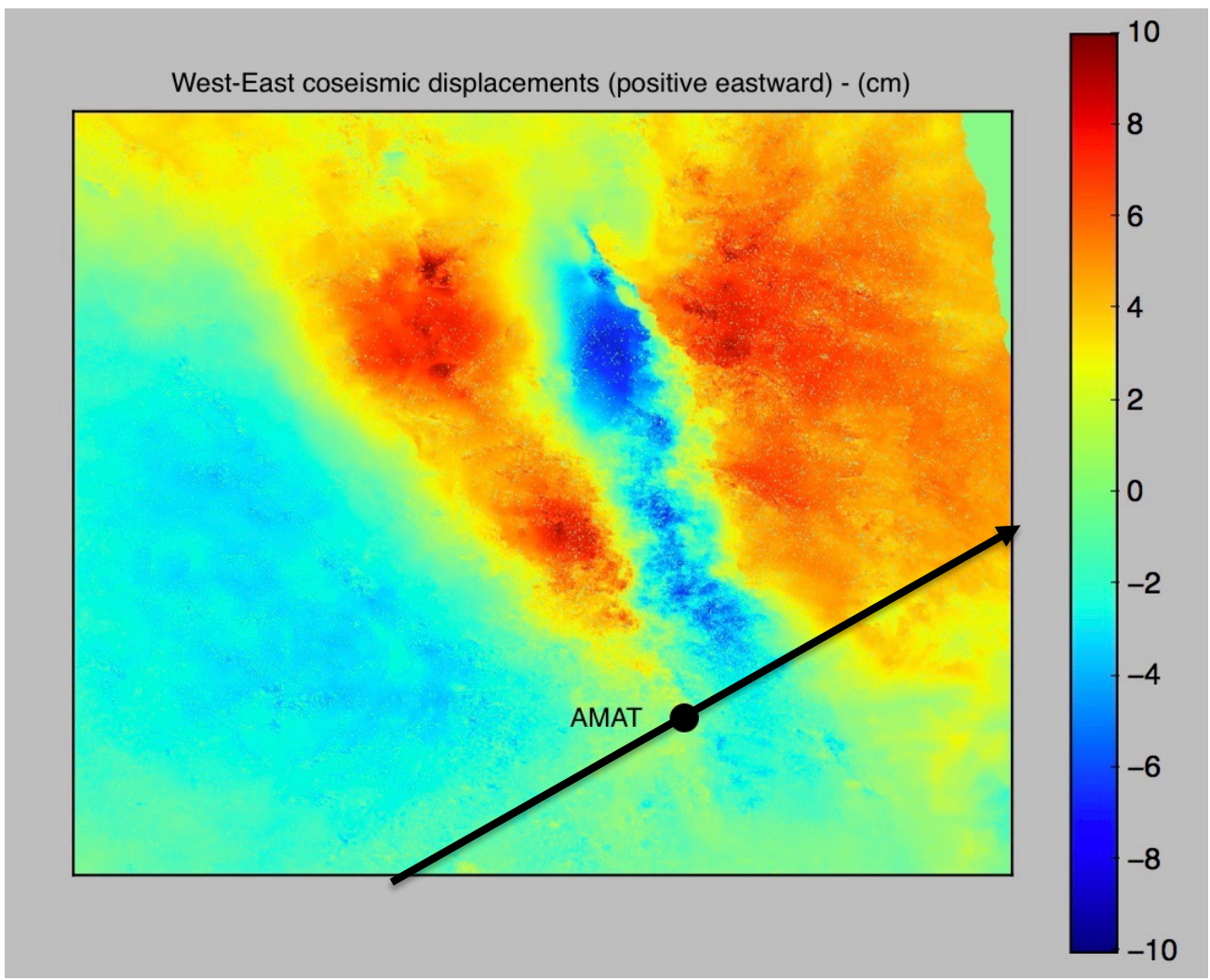

Figure 4 - West-East coseismic displacement field from SAR (as in Figure 2 - coloured for better understanding; the axis of the Amatrice transect is indicated by the black arrow).

The GPS data were further analyzed, in order to investigate if the mentioned peck velocity gradient in the area between Rieti and Amatrice changed significantly with time or not, during the whole time interval beginning of January 2005-middle of August 2016.

To this aim, at first, only the GPS stations active and reliable for about 2/3 of the whole time interval have been considered; this strict condition allowed for the selection of 14 GPS stations.

Then, a first analysis was performed arbitrarily splitting the whole time interval into two sub-intervals of almost equal lengths (beginning of January 2005-end of December 2010 and beginning of January 2011-middle of August 2016). In this case, the effect of the l'Aquila earthquake (April 6, 2009) on some GPS stations (jump in the positions time series) induces artefacts in their velocities estimation, albeit based on robust interpolation (Figure 5a). Therefore, a second analysis has been performed, splitting the whole interval into two sub-intervals, before and after the 2009 l'Aquila earthquake; in this case, no artefacts are seen (Figure 5b).

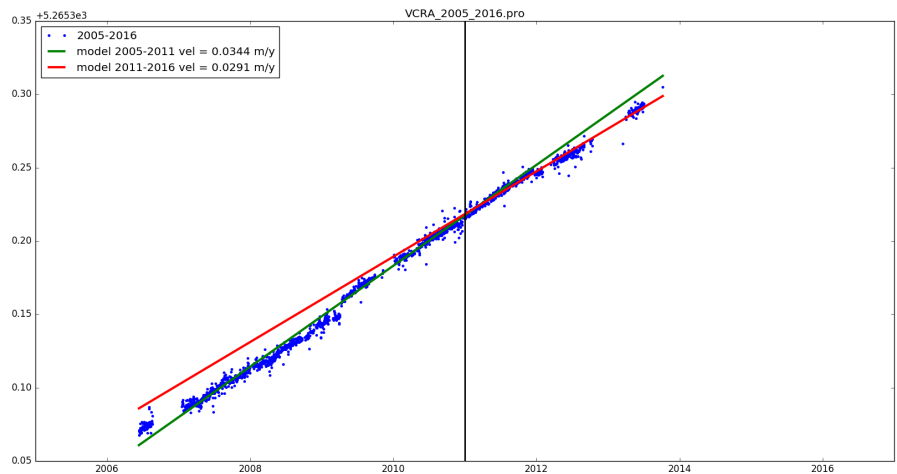

a)

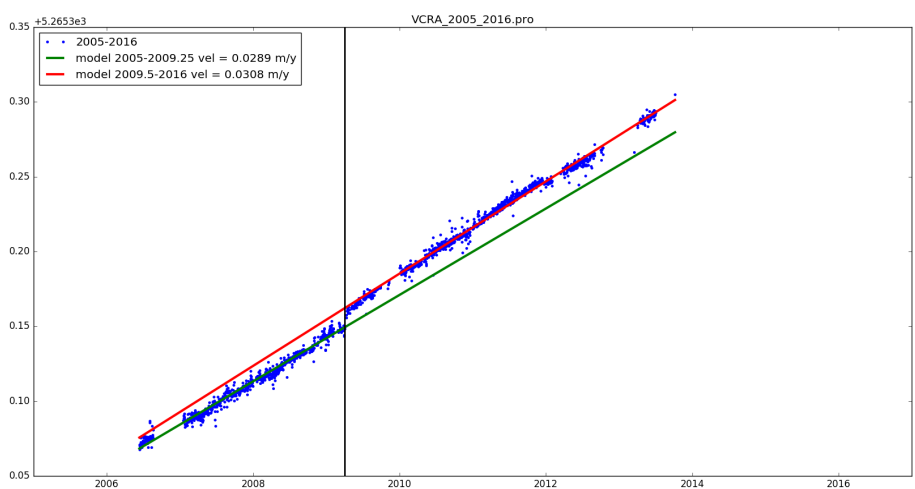

b)

Figure 5. GPS station VCRA – Linear robust velocities estimation along the Amatrice transect: a) before and after January 1st, 2011 (evident artefacts), b) before and after the l'Aquila 2009 earthquake (no evident artefacts).

A rather homogeneous behaviour before and after the l'Aquila earthquake has been assessed for all the selected 14 GPS stations along the Amatrice transect. No significant change with time has been detected for the velocity gradient in the area between Rieti and Amatrice, which is a long term pattern with respect to the time span of the available GPS data (Figure 6).

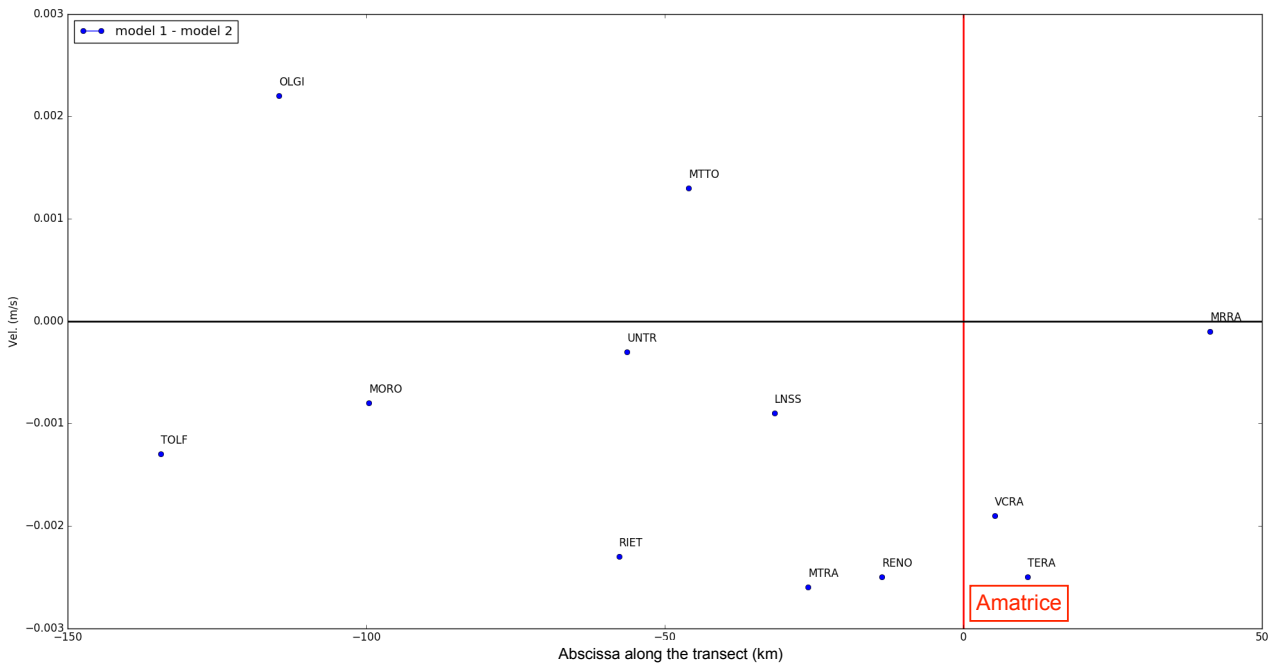

Figure 6 – Summary of the velocity differences before and after the 2009 l'Aquila earthquake for all the selected 14 GPS stations along the Amatrice transect.

**Additional analyses on the GPS velocity field: transects in Central and Northern Italy**

Similar analyses, based on GPS data only (Devoti et al., 2016), are carried on to investigate the velocity and strain pattern directed along:
- two transects 50 km wide and 180-200 km long in Central Italy, crossing the Apennines from the Tyrrhenian to the Adriatic sea along the direction of maximum tectonic extension (approximate azimuth 55°), with the axis passing respectively through Camerino (Figures 7 and 8) and Raiano (Figures 9 and 10); the analysis was approximately developed again for the time interval beginning of January 2005-middle of August 2016
- three transects 50 km wide and 130-160 km in Northern Italy, from the Apennines to the Po Valley along the direction of maximum tectonic shortening (approximate azimuth 25°), with the axis passing respectively through Finale Emilia (Figures 11 and 12), Imola (Figures 13 and 14) and Mantova (Figures 15 and 16); for these transects the GPS data were time limited (end of March 2012) before the M=6.0 Emilia earthquake occurred on May 20, 2012.

In spite of the fact that the number of GPS stations is remarkably lower than in the case of Amatrice transect, the other two transect across Central Italy and the three transects from the Apennines to the Po Valley supply quite important counter examples that remarkably support the validity of our findings in that they highlight, respectively:
- the overall tectonic extension across the Central Italy Apennines;

- the overall tectonic shortening moving northward from the Apennines to the Po Valley; the spatial deceleration increases moving eastward, passing from the Mantova to the Imola transect;
- the absence in the spatial acceleration of localized peaks with a trend comparable to the one well defined along the Amatrice transect between -50 and 0 km (Figure 3), well consistent with the fact that the CN alarm for the Emilia 2012 earthquake started on March 1st 2012, has a little overlap with the time range for which reliable geodetic data are available.

**Conclusions**

At first, it has to be underlined that an objective analysis, duly accounting for the alarm extents in space and time and the prediction failures, confirms the statistical significance of the CN results, which were able to predict the Amatrice earthquake.
Moreover, the integrated analysis of the available geodetic data coming from GPS and SAR indicates that it is possible to highlight (with this retrospective analysis) the velocity difference and the related strain accumulation in the area of Amatrice event, part of the CN alarmed area since November 1st, 2012 (DMG, 2017).
In the GPS data, the evidenced peak of the velocity gradient along the Amatrice transect does not show any change with time within the interval, beginning of January 2005-middle of August 2016. Therefore, the related strain accumulation appears to be a relatively long term phenomenon, even preceding CN alarm started November 2012.
The considered counter examples, i.e. transects (Camerino, Raiano, Mantova, Finale Emilia, Imola) across CN alarmed and non-alarmed areas (Figure 17), even if based on a relatively low density of data (roughly separated by 20 km, i.e. about half the wavelength of the mean velocity difference between the southern and northern parts of the transect boxed in Figure 3) do not show any peak velocity gradient comparable to the one well defined along the Amatrice transect.
Therefore, a denser and permanent GNSS network, possibly complemented by a continuous InSAR tracking (now possible thanks to the 6-days revisit time allowed by Sentinel-1A and 1B) may allow the routine highlight in advance of the velocity difference and the related strain accumulation and to cross correlate this analysis with the results of CN and other intermediate term middle range earthquake prediction algorithms, routinely already implemented. This integrated routine monitoring, CN and GPS, can be implemented in the near future, since, at present, dense permanent GNSS network can be established using low-cost GNSS receivers instead of high-cost geodetic ones.
An example of the possible significant reduction of the size of the CN alarmed areas, by the integrated monitoring, CN and GPS, is given in Figure 18. Here are reported the NDSHA ground motion scenarios at bedrock for alarmed Central Italy CN region. The scenarios are defined in terms of Peak Ground Velocity above 15cm/s that correspond to macroseismic intensity about $X_{MCS}$ (Indirli et al., 2011). Similar picture can be obtained considering DGA (Design Ground Acceleration) or PGA (Peak Ground Acceleration). The PGV values observed by Rete Accelerometrica Nazionale – Dipartimento Protezione Civile (RAN – DPC) at Amatrice (up to 31 cm/s) and Norcia (up to 56 cm/s) fit very nicely the values predicted by NDSHA ground motion scenarios (30-60 cm/s). The rectangle oriented as Amatrice transect delimits the area covered by the GPS stations reported in Figure 2. The intersection of the rectangle with the area covered by values of PGV≥15cm/s defines where preventive actions should be concentrated, accordingly with the joint monitoring.

# Acknowledgements

The Authors are indebted with Roberto Devoti , Federica Riguzzi, Riccardo Lanari, Manuela Bonano, Andrea Magrin and Volodya Kossobokov for contributing GPS data, SAR data and prior geological information, and for unselfish fruitful discussions that helped shaping up this paper.

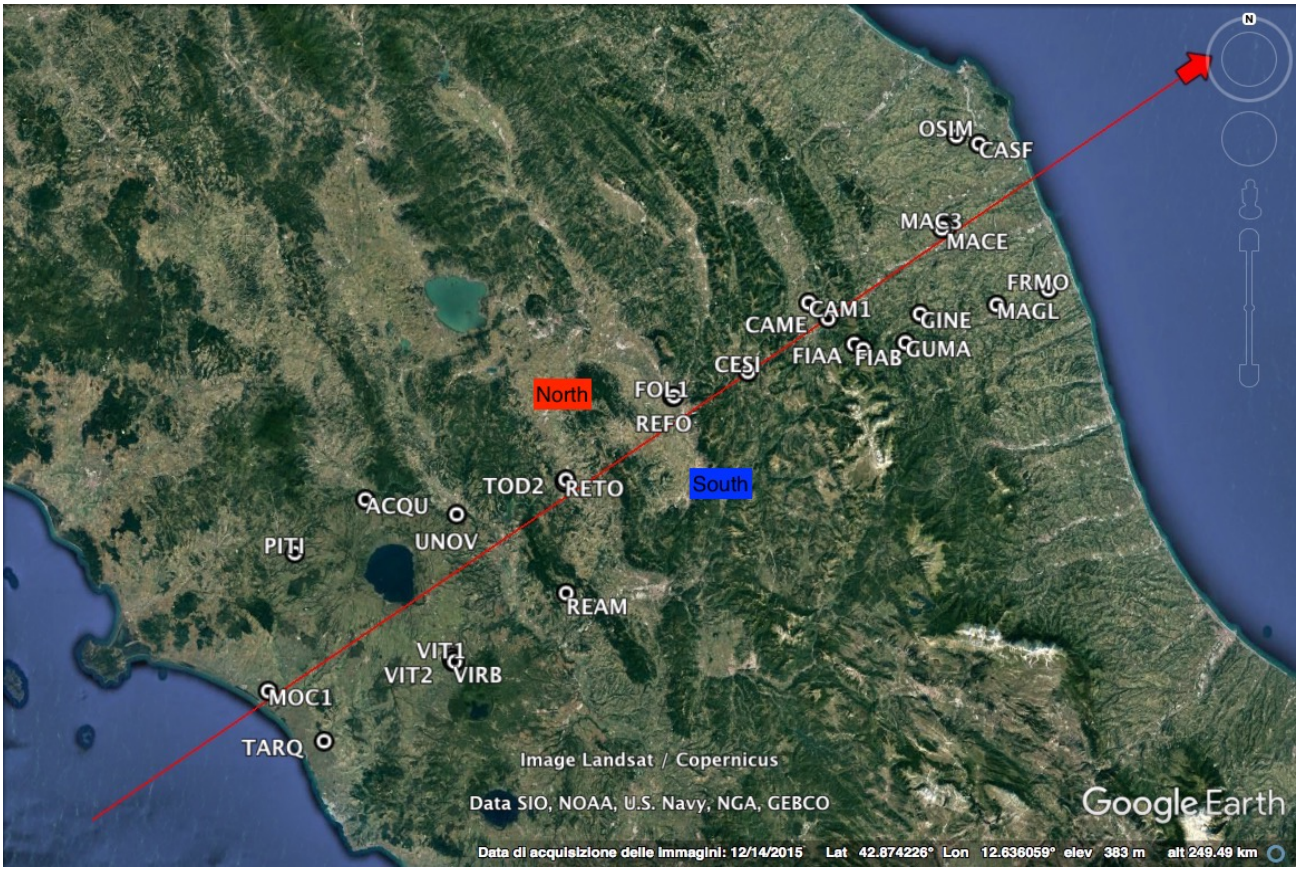

Figure 7 - Camerino transect across Apennines with GNSS stations.

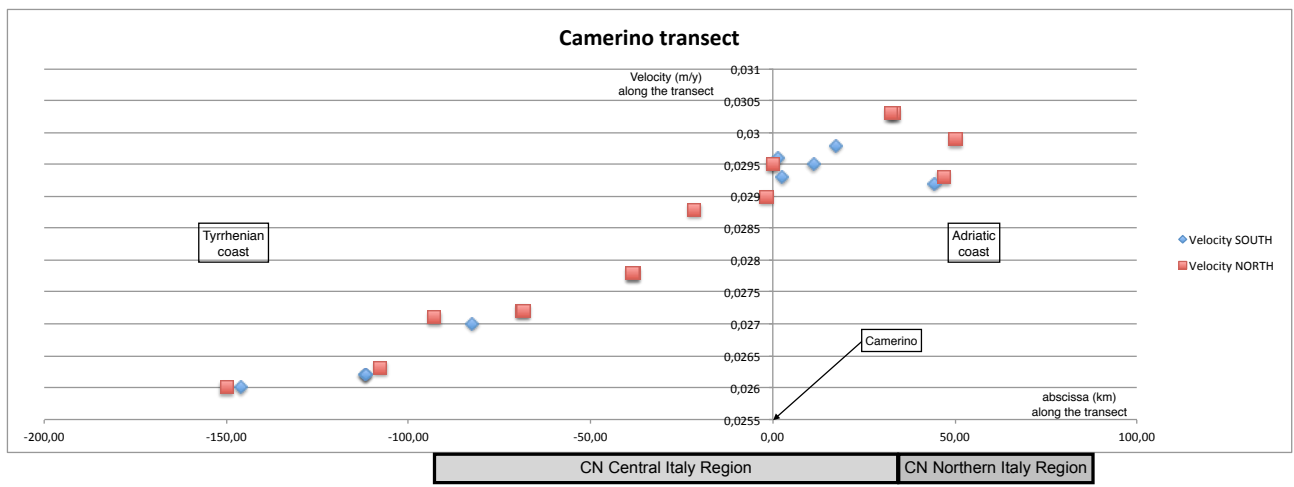

Figure 8 - Velocity pattern along the Camerino transect across Apennines (time range: beginning of January 2005 - middle of August 2016); Velocity South and North indicate the velocity patterns as derived by the GPS stations respectively located south and north with respect to the axis of the transect passing through Camerino (see Figure 7); the intersections of the axis of transect with the CN regions are indicated.

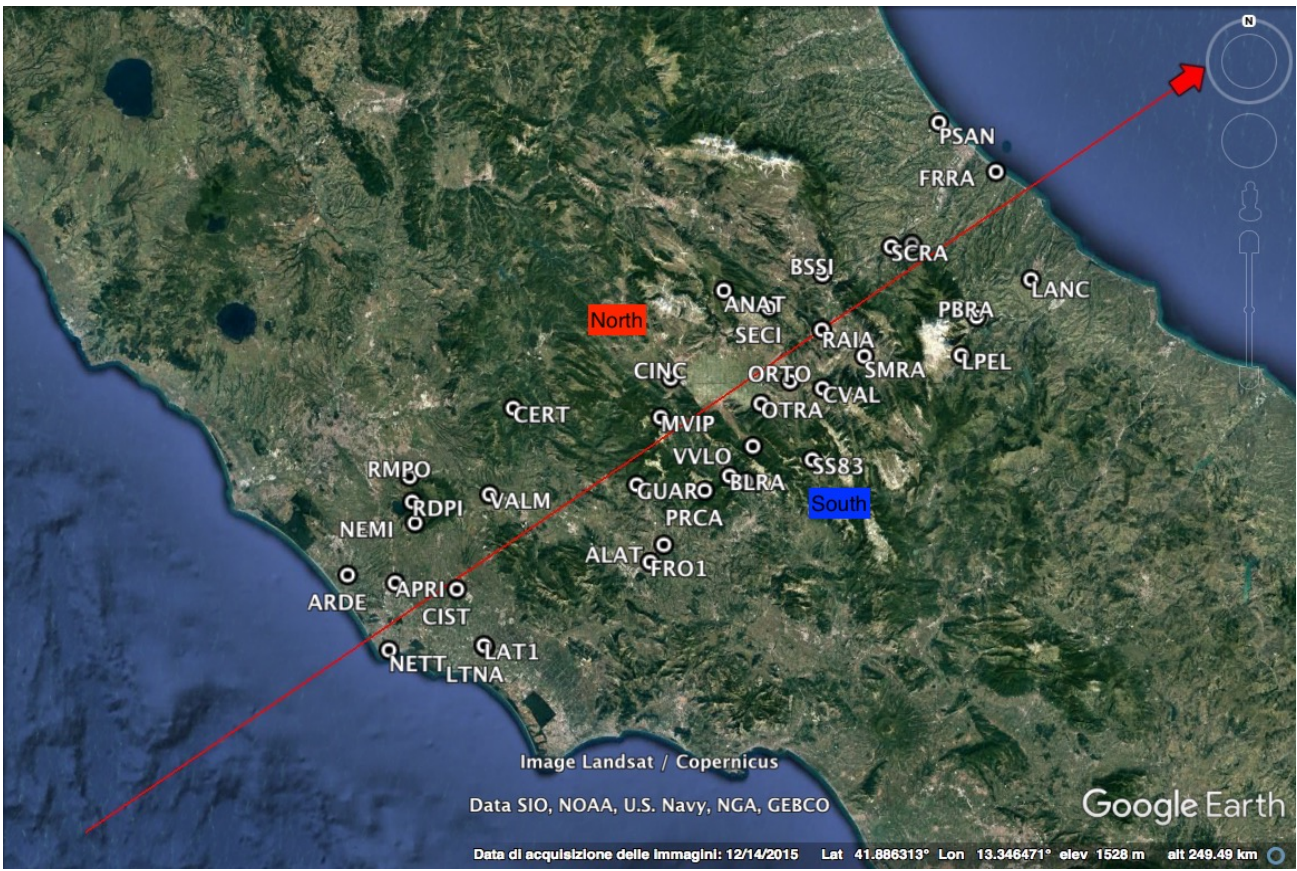

Figure 9 - Raiano transect across Apennines with GNSS stations.

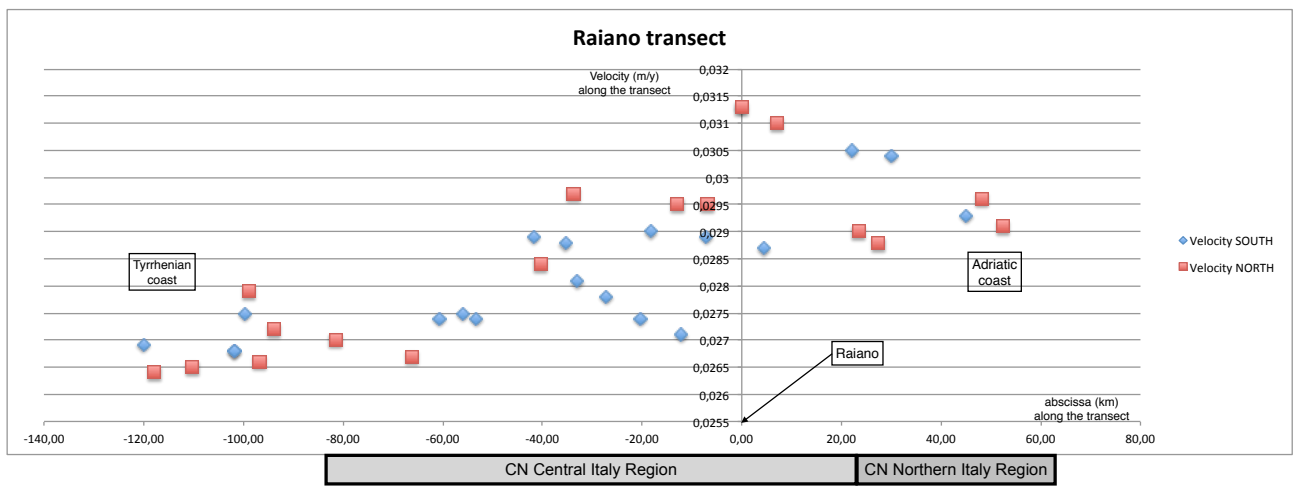

Figure 10 - Velocity pattern along the Raiano transect across Apennines (time range: beginning of January 2005 - middle of August 2016); Velocity South and North indicate the velocity patterns as derived by the GPS stations respectively located south and north with respect to the axis of the transect passing through Raiano (see Figure 9); the intersections of the axis of transect with the CN regions are indicated.

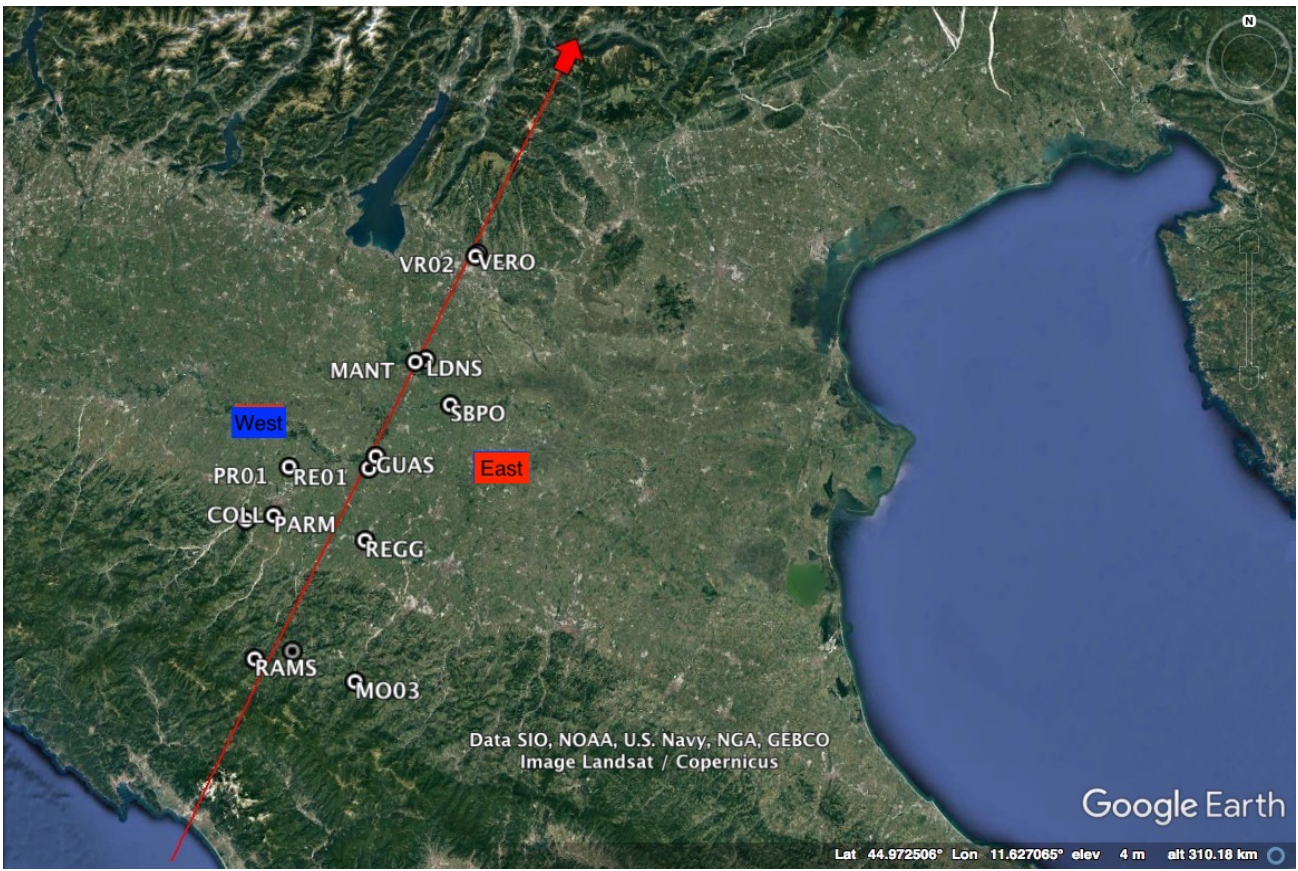

Figure 11 - Mantova transect from Apennines to Po Valley with GNSS stations.

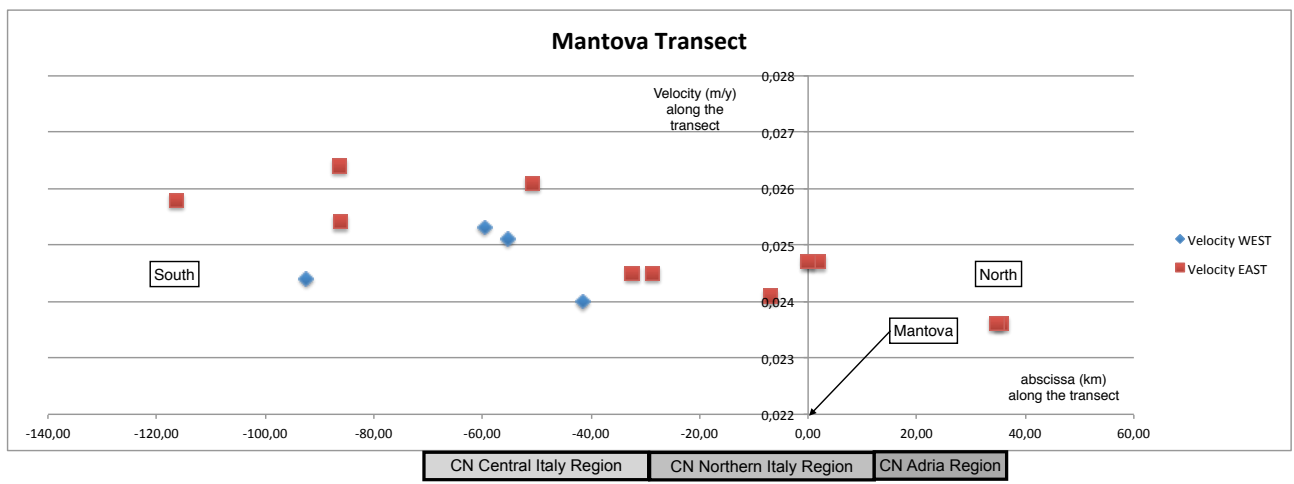

Figure 12 - Velocity pattern along the Mantova transect from Apennines to Po Valley (time range: end of May 2005 - end of March 2012); Velocity South and North indicate the velocity patterns as derived by the GPS stations respectively located south and north with respect to the axis of the transect passing through Mantova (see Figure 11); the intersections of the axis of transect with the CN regions are indicated.

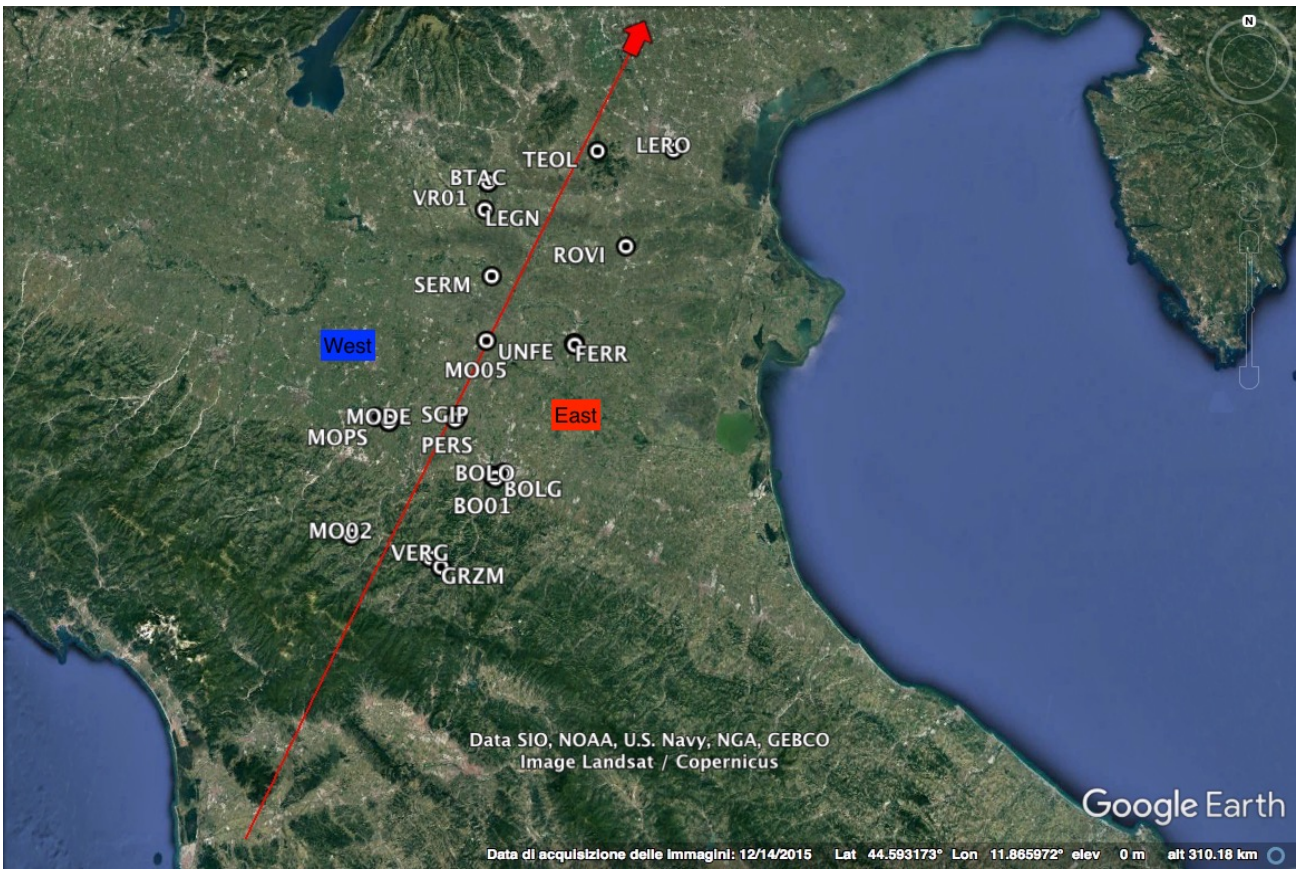

Figure 13 – Finale Emilia transect from Apennines to Po Valley with GNSS stations.

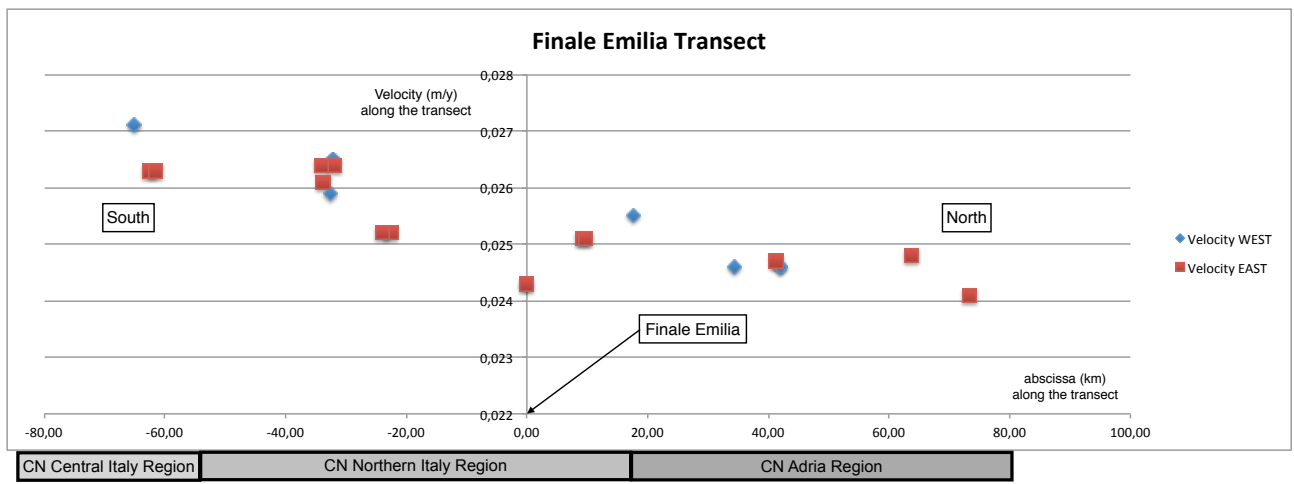

Figure 14 - Velocity pattern along the Finale Emilia transect from Apennines to Po Valley (time range: end of May 2005 - end of March 2012); Velocity South and North indicate the velocity patterns as derived by the GPS stations respectively located south and north with respect to the axis of the transect passing through Finale Emilia (see Figure 13); the intersections of the axis of transect with the CN regions are indicated.

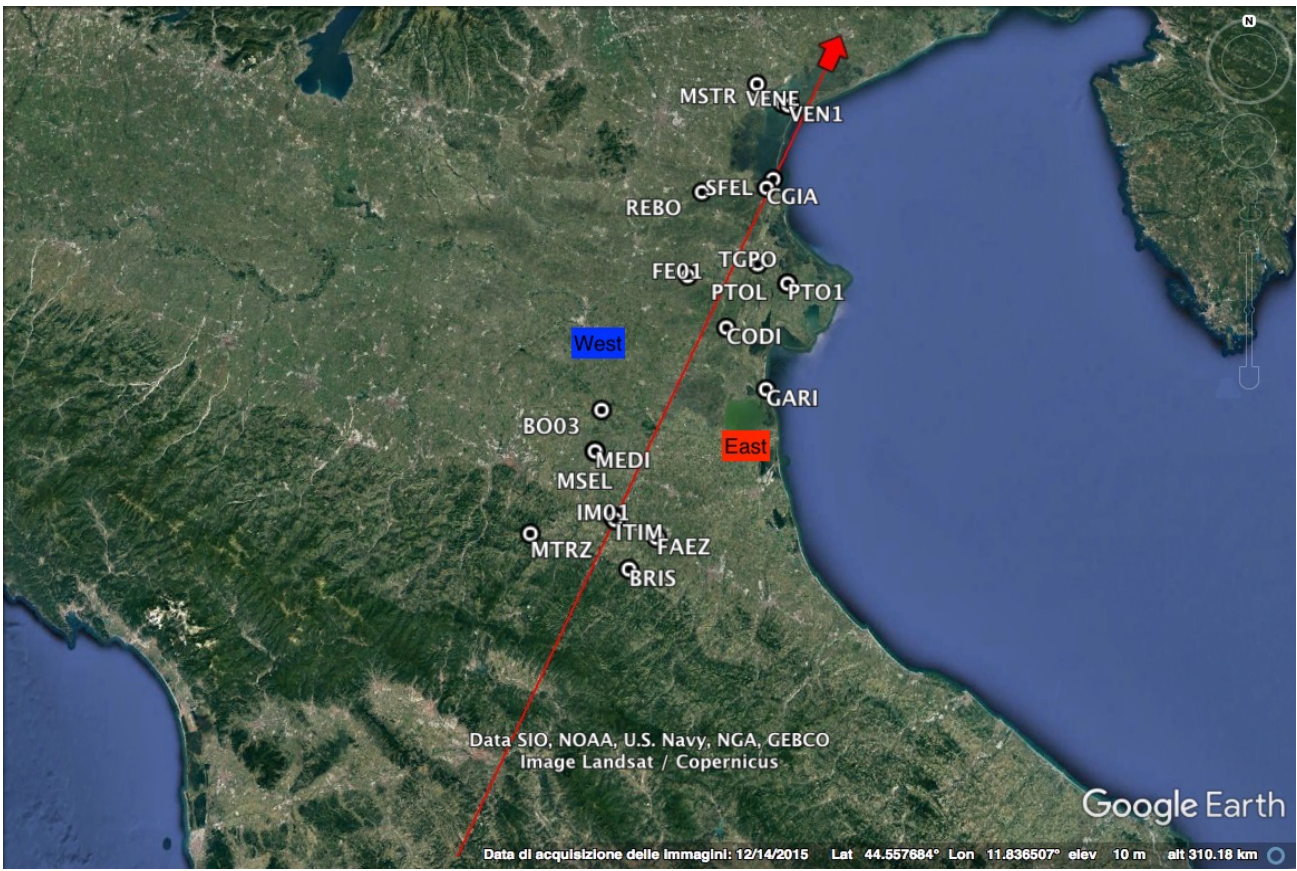

Figure 15 – Imola transect from Apennines to Po Valley with GNSS stations.

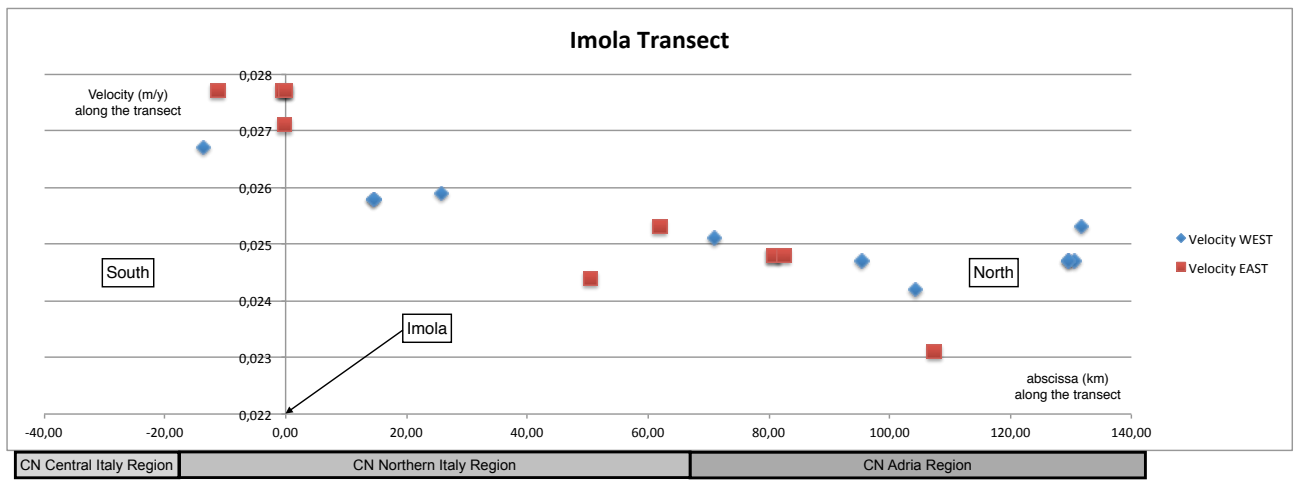

Figure 16 - Velocity pattern along the Imola transect from Apennines to Po Valley (time range: end of May 2005 - end of March 2012); Velocity South and North indicate the velocity patterns as derived by the GPS stations respectively located south and north with respect to the axis of the transect passing through Imola (see Figure 15); the intersections of the axis of transect with the CN regions are indicated.

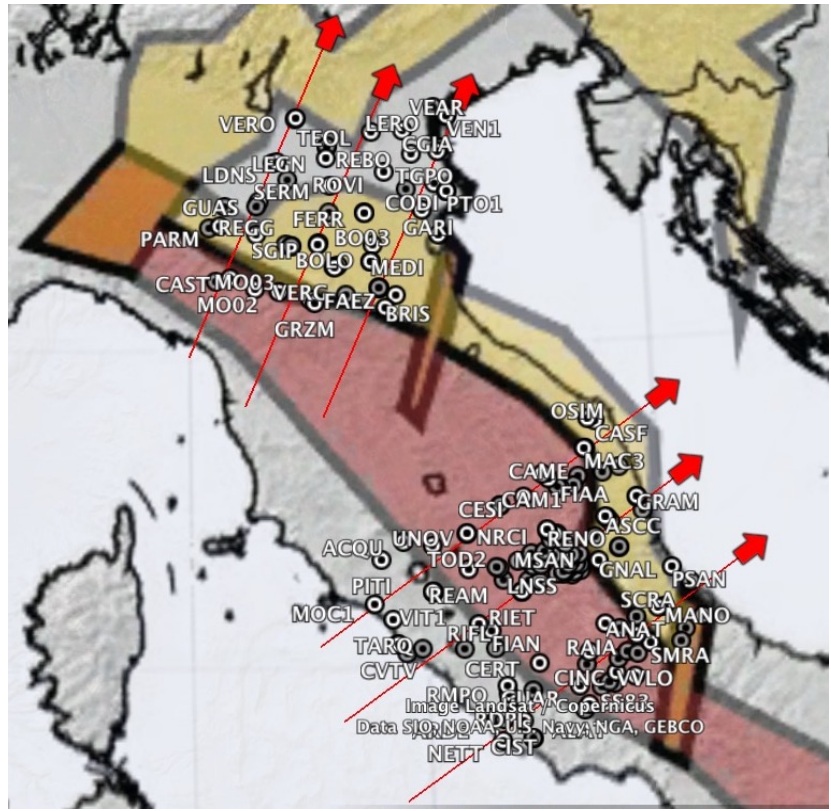

Figure 17 – CN Regions with the considered transects.

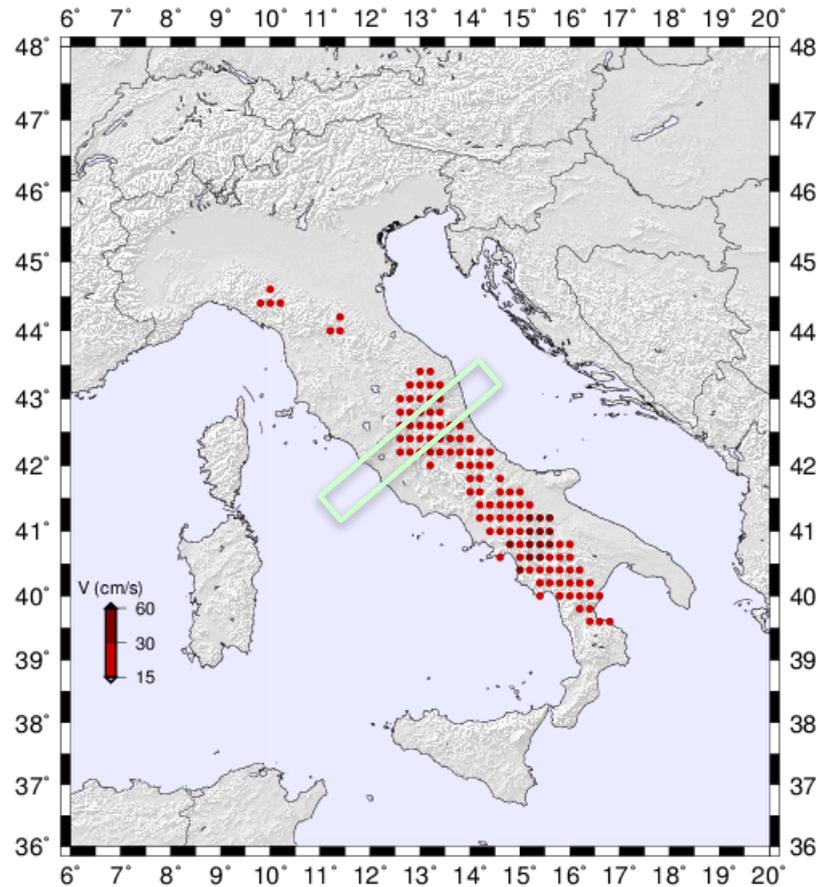

Figure 18 – Hazard scenario of CN Central Italy Region, expressed in terms of Peak Ground Velocity (PGV), and its intersection with the rectangular area covered by the RAN stations along the Amatrice transect.